\documentclass[twocolumn,amsmath,aps]{revtex4}
\usepackage{graphicx}

\newcommand{\nl}{\nonumber \\}

\newcommand{\cf}{cf.\ }

\newcommand{\be}{\begin{equation}}
\newcommand{\ee}{\end{equation}}
\newcommand{\bea}{\begin{eqnarray}}
\newcommand{\eea}{\end{eqnarray}}

\newcommand{\Eq}[1]{Eq.\,(\ref{#1})}

\newcommand{\la}{\langle}
\newcommand{\ra}{\rangle}
\newcommand{\dg}{\dagger}

\newcommand{\ti}{\tilde}

\begin{document}
\draft

\title{ Theory of Ultrafast Optical Manipulation of Electron Spins
        in Quantum Wells  }

\author{Jinshuang Jin and Xin-Qi Li}
\address{State Key Laboratory for Superlattices and Microstructures,
Institute of Semiconductors,
Chinese Academy of Sciences, P.O. Box 912, Beijing 100083, China  }

\date{\today}

\begin{abstract}
Based on a multi-particle-state stimulated Raman adiabatic passage
approach, a comprehensive theoretical study of the ultrafast
optical manipulation of electron spins in quantum wells is
presented. In addition to corroborating experimental findings
[Science {\bf 292}, 2458 (2001)], we improve the expression for the
optical-pulse-induced effective magnetic field, in comparison with
the one obtained via the conventional single-particle ac Stark
shift.
Further study of the effect of hole-spin relaxation
reveals that while the coherent optical manipulation
   of electron spin in undoped quantum wells
would deteriorate in the presence of
  relatively fast hole-spin relaxation,
  the coherent control in doped systems can be
   quite robust against decoherence.
The implications of the present results on
quantum dots will also be discussed.
\end{abstract}

\maketitle

\section{Introduction}

The electron spin, which has been largely ignored in
conventional charge-based electronics,
has now become a focus of research due to
the emerging field of spintronics \cite{Wol01,Aws99,Gup01a,%
Gup01b,Fuj02,Los00,Los03,Fol03,Li02,Pry02,Sha03,Pri9598}.
It is widely believed that incorporating the spin degree of freedom in
conventional charge-based electronics or using
electron spin alone as the information carrier will lead to
various applications in quantum devices. 
In particular, various spin-based quantum computer models have been proposed
\cite{Los98,Ima99,Los99,Pri98,Los01,Zol01,Fen03,Sol03}.

The development of spintronics largely depends on the ability
to control electron spins.
One of the most basic controls is coherent spin rotation,
which is traditionally implemented by making use of local magnetic fields.
However, in order to manipulate the desired coherent spin evolution,
the magnetic field pulses should in general
be shorter than the electron spin coherence time, which
is typically on the order of nanoseconds \cite{Cro95}.
Unfortunately, a sub-nanosecond magnetic field source
is technologically inaccessible at present.
Very recently, several theoretical studies have considered the possibility of
manipulating electron spin in solids by
{\it all-optical} means \cite{Li02,Pry02,Sha03}.
This approach has significant advantages since
the laser field is much more controllable than its
magnetic-field counterpart, and tunable femtosecond lasers
are now commercially available.
In fact, the ultrafast optical manipulation of electron spins
in quantum wells has been demonstrated experimentally \cite{Gup01a}.
The result has been understood
in terms of an effective magnetic field arising
from the ac Stark shift of
the single-electron state \cite{Gup01a,Ard61}.

In this paper, we present
a many-particle-state theory for the ultrafast optical
manipulation of electron spins in quantum wells
via the stimulated Raman adiabatic passage
(STIRAP) control scheme, in which
the effect of Pauli exclusion will be fully taken into account.
Furthermore, the consequences of hole-spin relaxation in the valence band
will be analyzed in detail,
for both an undoped system closely related to the experiment of
Ref.\ \onlinecite{Gup01a} and a doped system which has not yet
been experimentally investigated.
The remainder of this paper is organized as follows.
In Sec.\ II,  the concept of effective magnetic field,
which is crucial to the optical manipulation of electron spin,
is introduced first by considering the
ac Stark shift of single-particle state.
A many-particle STIRAP approach to coherent spin control
is then presented, together with
 numerical and perturbative results.
The effect of hole-spin relaxation is studied in Sec.\ III.
Doped systems will be considered in
Sec.\ IV,
where both the control of STIRAP manipulation and
the effect of hole-spin relaxation are studied
in comparison with their undoped counterparts
considered earlier.
Finally, concluding remarks are presented in Sec.\ V.

\section{Coherent Manipulation}

{\it ac Stark Effect and the Description of Effective Magnetic Field}.---
The basic idea of an optical approach to
manipulating electron spin is to make use of an off-resonance
laser pulse to induce ac Stark shifts, which are in turn
equivalent to an effective magnetic field
\cite{coh72,com88,jof89}.
This idea has recently been employed to manipulate electron spin
in undoped semiconductor quantum wells \cite{Gup01a}. Initially,
an electron is pumped from the valence band to the conduction band
by a resonant laser pulse along the $z$-direction and with
$\sigma^+$-polarization. The resultant state diagram
is shown in Fig.\ 1(a), where all the states are in the
``$z$"-representation \cite{note-zx}.
In the absence of magnetic field, the lowest conduction-band (CB)
level is two-fold degenerate, denoted by spin states
$|\pm 1/2\rangle_c$; and the valence-band (VB) states are denoted
by $|\pm3/2\rangle_v$ and $|\pm1/2\rangle_v$. Here,
we have taken into account the splitting of the valence band-edge
states from the original four-fold degenerate states into two two-fold
degenerate states, due to the breakdown of the symmetry of the quantum well.
Further consideration of the exciton effect
leads to the shifted electron-hole transition energies,
$E_{X_1}=(E_c-E_{v_1})-U_{X_1}$
and $E_{X_2}=(E_c-E_{v_2})-U_{X_2}$.
Here $E_c$, $E_{v_1}$ and $E_{v_2}$ denote the
energies of $|\pm 1/2\rangle_c$,
$|\pm3/2\rangle_v$ and $|\pm1/2\rangle_v$, respectively, and
$U_{X_1}$ and $U_{X_2}$ are the Coulomb interaction
(attraction) energies of the corresponding electron-hole pairs.
The discrete-level model of Fig.\ 1(a) for a quantum well is based on
two considerations: (i) the initial preparation of the electron-hole
pair is accomplished by the resonant excitation between the
band-edge states; (ii) the subsequent off-resonant optical
manipulation involves only
the band-edge states because of
the negligibly small momenta of the photons
causing only vertical (and also virtual) transitions.
Therefore, states in the diagram shown in Fig.\ 1(a) should be understood
as the band-edge states.

\begin{figure}\label{Fig1}
\includegraphics*[scale=0.5,angle=0.]{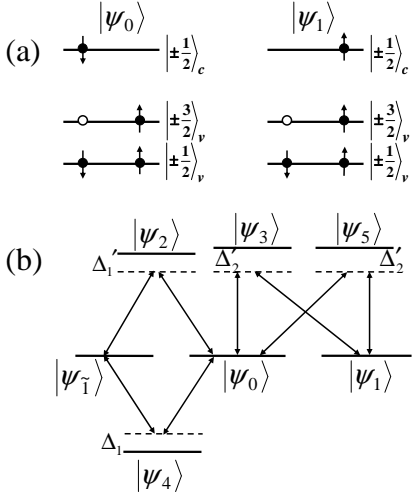}
\caption{Diagram of band-edge states of undoped quantum well after
optically exciting an electron from the valence to the conduction
band. (a) Two degenerate state configurations with different
electron spin in the conduction band, between which quantum coherent
oscillation is to be performed. (b) Intermediate states which
mediate transition between $|\psi_0\ra$ and $|\psi_1\ra$ during the
optical manipulation. }
\end{figure}

To manipulate the electron spin in the conduction band, let us apply
a below-band-gap (negative-detuning) laser pulse of the same
$\sigma^{+}$-polarization but along the $x$-direction. This laser
induces $ |-1/2\ra_{c,x} \leftrightarrow |-3/2\ra_{v,x} $ and $
|1/2\ra_{c,x} \leftrightarrow |-1/2\ra_{v,x} $ transition couplings
between the specified states in the ``$x$''-representation.
Other 
transitions
are forbidden due to the selection rule.
It is well known that this kind of off-resonance coupling will cause
ac Stark shifts of the relevant energy levels. In particular,
from second-order perturbation theory the energy shifts of the CB
states $|-1/2\rangle_{c,x}$ and $|1/2\rangle_{c,x}$
can be estimated as \cite{Li02}:
\begin{subequations}\label{shift}
\bea
\Delta E_{-1/2}^{c,x}&=&|V_{-1/2,-3/2}|^{2}/\Delta_{1}, \\
\Delta E_{1/2}^{c,x}&=|&V_{1/2,-1/2}|^{2}/\Delta_{2}.
\eea
\end{subequations}
Here, $\Delta_{1}=E_{X_1}-\hbar\omega_p$
and $\Delta_{2}=E_{X_2}-\hbar\omega_p$ are
the detunings of the photon energy ($\hbar\omega_p$)
with respect to the excitation energies
from the heavy and light valence band states, respectively,
to the conduction band state.
$V_{-1/2,-3/2} $ and $V_{1/2,-1/2}$ are the coupling matrix
elements of the laser with the electron-hole pair states.
Explicitly, from energy-band theory \cite{Bas88},
the spatial wave-functions of the considered states are composed of
the Bloch functions $|S\ra$, $|X\ra$, $|Y\ra$ and $|Z\ra$.
Denoting $\Omega\equiv eE_0\la S|x|X\ra=eE_0\la S|y|Y\ra=eE_0\la S|z|Z\ra$,
the coupling matrix elements read
\begin{subequations}\label{Veh}
\bea
&& V_{-1/2,-3/2}=eE_{0} {_{c,x}}\langle -1/2|\vec{r}\cdot
\vec{\epsilon}|-3/2\rangle_{v,x}      \nl
&&~~~~=-ieE_{0}\langle S|y+iz|Y-iZ\rangle/\sqrt{2}=-i\sqrt{2}\Omega,  \\
&& V_{1/2,-1/2}=eE_{0}{_{c,x}}\langle 1/2|\vec{r}\cdot
\vec{\epsilon}|-1/2\rangle_{v,x}                \nl
&&~~~~=ieE_{0}\langle S|y+iz|Y-iZ\rangle/\sqrt{6}=i\sqrt{2/3}\Omega.
\eea
\end{subequations}
Here the equation $\vec{r}\cdot\vec{\epsilon}=y+iz$ stems from
the $\sigma^+$-polarization and $x$-direction propagation of the laser pulse.

In practice, the ac Stark shifts can be conveniently described as
an effective magnetic field $B_{eff}$ along the manipulating laser direction.
In terms of the CB level splitting
$\delta_{CB}=\Delta E_{-1/2} ^{c,x}-\Delta E_{1/2}^{c,x}$,
the effective magnetic field reads
\begin{equation}\label{Beff}
B_{eff}=\delta_{CB}/(g_{e} \mu_{B}).
\end{equation}
Here $g_{e}$ is the Land$\acute{e}$-$g$ factor and $\mu_{B}$ is
the Bohr magneton. In addition, the duration of this
effective magnetic field is equivalent to the duration of the laser-pulse, which
can be as short as femtoseconds.
This effective magnetic field description
can be further elaborated as follows.
The initially created CB electron spin state
$|1/2\ra_{c}$ is an eigenstate of the Pauli operator $\sigma_z$.
It can be expressed in terms of the eigenstates of $\sigma_x$ as
$|\Psi(t=0)\rangle=|1/2\ra_{c}
=\frac{1}{\sqrt{2}}\left(|1/2\ra_{c,x}+|-1/2\ra_{c,x}\right)$.
Upon switching on the off-resonance laser pulse along the
$x$-direction, the ac Stark shifts $\Delta E_{\pm 1/2} ^{c,x}$
would result in a relative phase difference $\phi=(\Delta
E_{1/2}^{c,x}-\Delta E_{-1/2} ^{c,x})t$, between the states
$|1/2\ra_{c,x}$ and $|-1/2\ra_{c,x}$.
Recasting $|\Psi(t)\ra$ in the $\sigma_z$-eigenstate representation
leads to $|\Psi(t)\rangle=\cos\frac{\phi}{2}|1/2\ra_{c}
+i\sin\frac{\phi}{2}|-1/2\ra_{c}$.
Clearly, the off-resonant laser pulse along the $x$-direction
plays the role of an effective magnetic field pulse
with magnitude determined by \Eq{Beff}.


{\it Many-Particle STIRAP Approach}.--- In the following we present
an improved effective magnetic field description
based on a many-particle state STIRAP
approach. This approach is a direct generalization of the
single-particle STIRAP in quantum optics \cite{Scu97} that takes into
account the multi-electron occupation of the valence band. The
effective coupling between $|1/2\ra_{c}$ and $|-1/2\ra_{c}$ will
then be established via a number of intermediate many-electron
states as depicted in Fig.\ 1(b).

For convenience, we introduce Fock's
particle-number representation to denote the multi-electron state.
For instance, we denote
$|\psi_{0}\rangle\equiv|1,0;0,1,1,1\rangle$, and
$|\psi_{1}\rangle\equiv|0,1;0,1,1,1\rangle$, where ``1" (``0")
stands for the occupation (vacancy) of the individual single
particle states, which correspond to, respectively, the CB states
$|\mp 1/2\ra_c$, and the VB states $|\mp 3/2\ra_v$ (heavy holes)
and $|\mp 1/2\ra_v$ (light holes).
Under the action of the off-resonance laser pulse described previously,
the CB states will be virtually coupled to the VB states.
As a consequence,
a number of intermediate virtual states establish an effective interaction
between $|\psi_{0}\ra$ and $|\psi_{1}\ra$.
Depending on the polarization of the laser pulse, selection rules
only allow the following intermediate states \cite{Bas88,Not-1}:
$|\psi_{2}\rangle\equiv|0,0;1,1,1,1\rangle$,
$|\psi_{3}\rangle\equiv|1,1;0,1,1,0\rangle$,
$|\psi_{4}\rangle\equiv|1,1;0,0,1,1\rangle$,
and $|\psi_{5}\rangle\equiv|1,1;0,1,0,1\rangle$.
Also, based on the selection rule, a possible state,
$|\psi_{\tilde{1}}\rangle\equiv|0,1;1,0,1,1\rangle$,
may be correlated with the initial state $|\psi_{0}\rangle$.
All these relevant multi-electron states are shown in Fig.\ 1(b).
In the absence of spin relaxation, the subspace spanned by these basis states
is complete for the evolution of the system.
We refer to this subspace as {\it coherent subspace},
and denote it by $\textbf{M}^{coh}=\{|\psi_{i}\rangle,i=0,1,\tilde{1},2,3,4,5\}$.
In $\textbf{M}^{coh}$ the driving Hamiltonian (in the interaction picture) reads
\begin{equation}\label{Hami}
 H=\left( \begin{array}{ccccccc}
  0 & 0 & 0 & \Omega_{20}^{\ast} & \Omega_{30}^{\ast} & \Omega_{40}^{\ast} & \Omega_{50}^{\ast}
  \\
  0 & 0 & 0 & 0 & \Omega_{31}^{\ast} & 0 & \Omega_{51}^{\ast} \\
  0 & 0 & 0 & \Omega_{2\tilde{1}}^{\ast} & 0 & \Omega_{4\tilde{1}}^{\ast} & 0 \\
  \Omega_{20} & 0 & \Omega_{2\tilde{1}} & -\Delta_{1} & 0 & 0 & 0 \\
  \Omega_{30} & \Omega_{31} & 0 & 0 & \Delta'_{2} & 0 & 0 \\
  \Omega_{40}& 0 & \Omega_{4\tilde{1}} & 0 & 0 & \Delta'_{1} & 0 \\
  \Omega_{50} & \Omega_{51} & 0 & 0 & 0 & 0 & \Delta'_{2} \\
\end{array}
\right).
\end{equation}
Here $\Delta'_1=\Delta_1+U_{XX}$, and $\Delta'_2=\Delta_2+U_{XX}$,
where $U_{XX}$ is the Coulomb repulsive energy
of two excitons that appear in the intermediate virtual states.
In \Eq{Hami},
$\Omega_{ij}=eE_{0}\langle\psi_{i}|\vec{r}\cdot\vec{\epsilon}|\psi_{j}\rangle$
describes the laser-induced coupling between the conduction and
valence band states.
Similar to the calculation in \Eq{Veh},
$\Omega_{ij}$ can be straightforwardly determined by using
\begin{subequations}
\bea
eE_{0}\langle S|y+iz|X+iY\rangle &=& i\Omega,
\\
eE_{0}\langle S|y+iz|X-iY\rangle &=& -i\Omega,
\\
eE_{0}\langle S|y+iz|Z\rangle &=& i\Omega .
\eea
\end{subequations}
Note that here all the states are expressed in the ``$z$"-representation,
while the states in \Eq{Veh} are in the ``$x$"-representation.
Explicitly, we obtain
$\Omega_{20}=\Omega_{4\tilde{1}}=-\Omega_{40}
=-\Omega_{2\tilde{1}}=-\Omega/\sqrt{2}$,
$\Omega_{50}=\Omega_{31}=\Omega/\sqrt{6}$,
$\Omega_{30}=\Omega_{51}=-\sqrt{2/3}\Omega$.

\begin{figure}\label{Fig2}
\includegraphics*[scale=0.5,angle=0.]{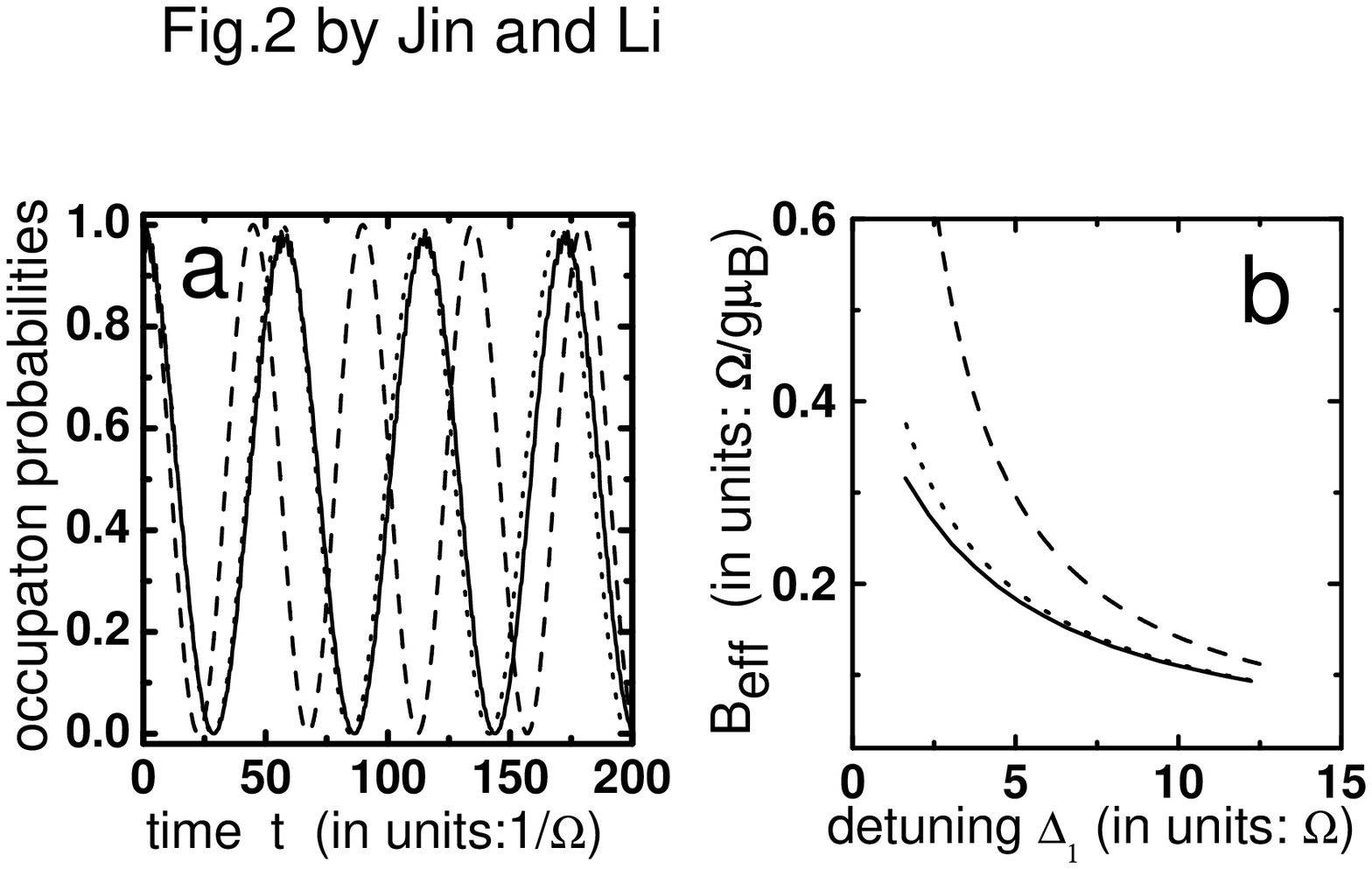}
\caption{ Coherent manipulation of electron spin in the absence of
hole spin relaxation in the valence band. (a) Laser-induced Rabi
oscillation between the spin-up and spin-down states, where the
results depicted by the solid, dotted and dashed curves are
respectively from three approaches: the direct numerical calculation
with the coherent subspace Hamiltonian of \Eq{Hami}, the effective
two-state multi-particle STIRAP Hamiltonian of \Eq{Heff} or
(\ref{Beff-2}), and the analysis based on the single-particle-state
picture of \Eq{Beff}. (b) The corresponding effective magnetic
fields extracted from these three methods. Here $\Delta_{1}$ denotes
the detuning. }
\end{figure}

In analogy to the (three-state) $\Lambda$ atomic system
in quantum optics \cite{Scu97},
the problem under study
can now be solved via a many-particle multi-state STIRAP approach.
Before giving a numerical demonstration,
let us consider an effective many-particle
two-state Hamiltonian approach to
the description of
the STIRAP coupling between $|\psi_0\ra$ and $|\psi_1\ra$.
Notice that $|\psi_{0}\rangle$ also virtually
couples to $|\psi_{\tilde{1}}\rangle$ through two paths  [\cf Fig.\ 1(b)]
via the intermediate states $|\psi_{2}\rangle$
and $|\psi_{4}\rangle$, respectively.
However, their opposite excitation configurations lead to
cancellation of these two paths due to destructive interference,
provided that the magnitude of the detuning is much larger than the exciton
interaction energy $U_{XX}$. The above analysis will be verified
numerically soon.
Therefore, we may only need to consider the two paths connecting
$|\psi_{0}\rangle$ and $|\psi_{1}\rangle$
via  $|\psi_3\ra$ and $|\psi_5\ra$.
Adiabatically eliminating the intermediate states leads to the effective
two-state multi-particle STIRAP Hamiltonian \cite{Scu97}
\begin{subequations}\label{Heff}
\begin{equation}
H_{eff}= -\tilde{\Omega}(|\psi_{0}\rangle\langle\psi_{1}|
        +|\psi_{1}\rangle\langle\psi_{0}|)/2,
\end{equation}
with the Rabi frequency
\begin{equation}
\tilde{\Omega}/2=|\Omega_{31}\Omega_{30}
+\Omega_{51}\Omega_{50}|/\Delta'_2.
\end{equation}
\end{subequations}
By identifying the Rabi frequency with the Larmor frequency
of spin precession around a magnetic field,
the {\it effective magnetic field} can be defined as
\bea\label{Beff-2}
\ti{B}_{eff}=\ti{\Omega}/(g_e\mu_B) .
\eea
Remarkably, the following numerical calculation will demonstrate
that this is a considerable improvement over \Eq{Beff},
as a result of taking into account the multi-electron occupation in the valence band.

{\it Numerical Results}.---
 In accordance with the experiment of Ref.\ \onlinecite{Gup01a},
where the spin rotation period under optical manipulation is on a
timescale of picoseconds, we choose the following parameter values
in our numerical study: the Rabi frequency $\Omega= 4$ meV,
the detuning parameters $\Delta_{1}=10\,\Omega$ and $\Delta_{2}=11.5\,\Omega$,
and the exciton Coulomb interaction energy $U_{XX}=0.5\,\Omega$.
Figure 2(a) depicts the laser induced Rabi oscillations between the spin-up
and spin-down states of the CB electron.
We present results from three approaches: the direct numerical
calculation with the coherent subspace Hamiltonian of \Eq{Hami},
the effective two-state multi-particle STIRAP Hamiltonian of \Eq{Heff}
or (\ref{Beff-2}),
and the analysis based on the single-particle-state picture of \Eq{Beff}.
We find that the effective two-state multi-particle STIRAP Hamiltonian
approach gives a result almost identical
with the exact numerical calculation,
while the result based on the single-particle ac Stark
shift differs considerably.
As we have already analyzed, the Rabi oscillation of spin can be equivalently
described as a the Larmor precession around an effective magnetic field.
In Fig.\ 2(b) we show the effective magnetic field
complementary to the Rabi oscillation in Fig.\ 2(a).
In the adiabatic regime (i.e.\ the large detuning case),
the multi-particle-state-based effective magnetic
field given by \Eq{Beff-2} can precisely describe
the CB electron spin rotations.
In contrast, the single-particle-state-based counterpart given
by \Eq{Beff} is inaccurate and will cause considerable errors in practice.

So far, we have only considered the rotation of the conduction electron spin.
Now we briefly discuss the possible effects of
hole-spin rotation, which has been typically ignored in the analysis of
Faraday rotation experiments \cite{Gup01a,Gup01b}
because of its rapid relaxation with respect to
the conduction electron spin.
In general, the hole-spin rotation should also affect
the Faraday rotation signals
in the coherent (or short-time) regime.
Interestingly, it can be shown that
for the specific optical excitation-manipulation setup
studied in this work,
the off-resonant laser along the $x$-direction
rotates only the conduction electron spin
but does not affect the hole spin at all.
Of course, both the electron and hole spins would be rotated
if the manipulating laser propagates along a different direction.
Probing both the electron and hole spin rotations should
be an interesting subject of Faraday
rotation experiments in general.

\section{Effect of Hole Spin Relaxation}

In reality, both the CB electron spin and VB hole spin
will suffer environment-induced scattering, and have finite decoherence times.
In Ref.\ \onlinecite{Sha03},
electron-hole recombination was considered
the dominant source of decoherence for quantum dots.
For the quantum well studied in this work, however,
the dominant source of decoherence should be
the VB hole-spin relaxation. It typically occurs in a matter of
picoseconds \cite{Gup01a,Gup01b,Uen90,Rou92},
much shorter than the typical timescale of nanoseconds
for electron-hole recombination
and CB electron spin relaxation \cite{Gup01a,Gup01b,Cro95}.

The hole-spin relaxation involves the following
incoherent VB state jumps:
$|3/2\rangle_v \leftrightarrow |-3/2\rangle_v$,
$|1/2\rangle_v \leftrightarrow |-1/2\rangle_v$, and
$|\pm 3/2\rangle_v \leftrightarrow |\pm 1/2\rangle_v$.
These processes can be described
with the spin-jump operators
$S_{1}=|-3/2\rangle_v \langle3/2|$, $S_{2}=|-1/2\rangle_v \langle1/2|$,
and $S_{3,4,5,6}=|\pm1/2\rangle_v\langle\pm3/2|$,
respectively,
together with their Hermitian conjugates.
\begin{figure}\label{Fig3}
\includegraphics*[scale=0.5,angle=0.]{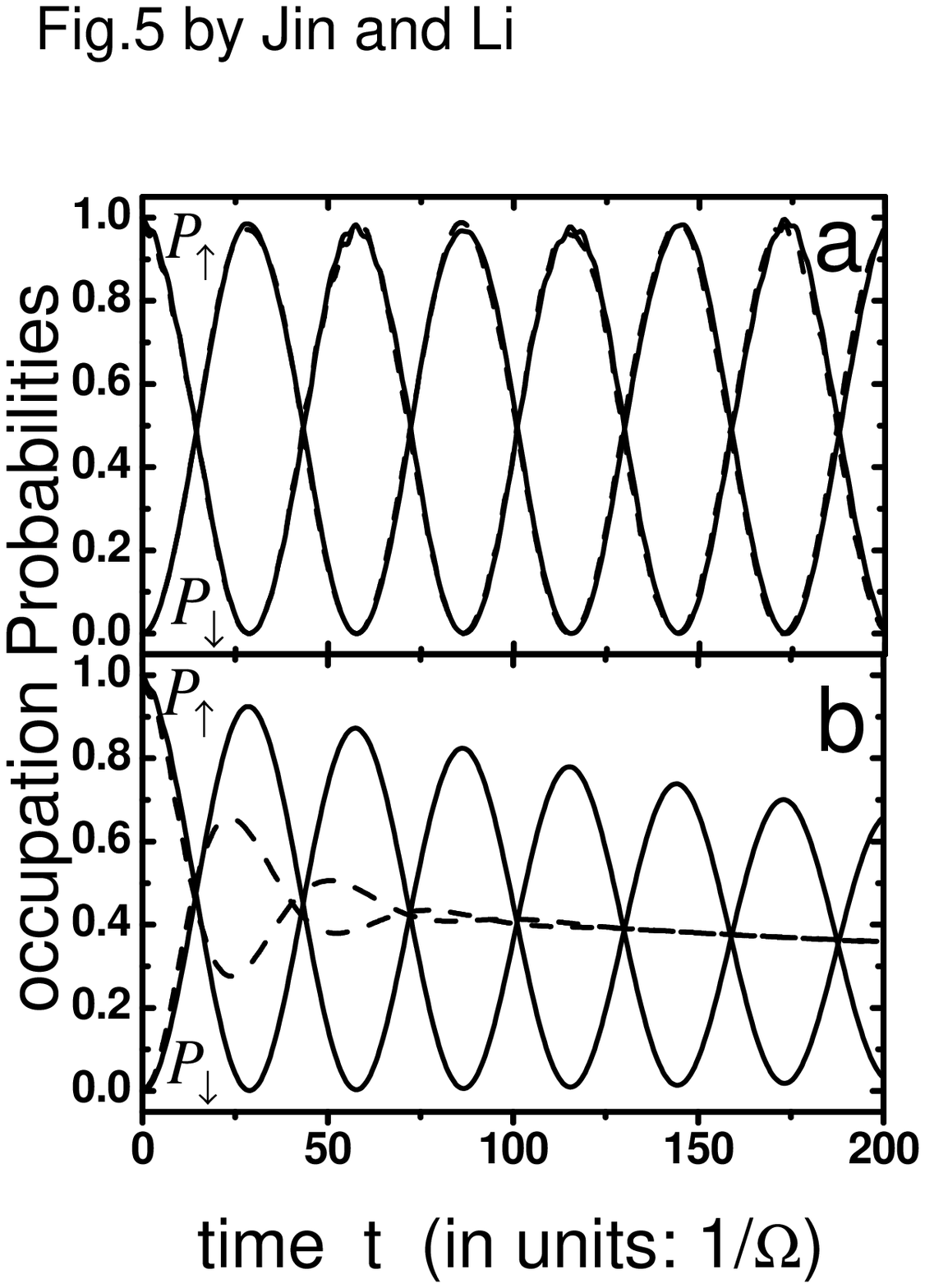}
\caption{ Hole spin relaxation effect on the manipulation of the
conduction-band electron spin, in the absence (a) and presence [(b)
and (c)] of pre-relaxation of the hole spin, respectively. }
\end{figure}
In particular, we employ the well-known Lindblad master equation to address
the hole spin relaxation effect, which reads \cite{Scu97,Lin76}
\begin{equation}\label{Lindblad}
\dot{\rho}=-i[H,\rho]-\sum\limits_{j}\gamma_{j}D[S_{j}]
-\sum\limits_{j}\tilde{\gamma}_{j}D[S_{j} ^{\dg}].
\end{equation}
The Lindblad superoperators are
$D[S_{j}]=\frac{1}{2}\{S^{\dg} _{j}S_{j},\rho\}-S_{j}\rho S_{j}^{\dg}$
and $D[S^{\dg}_{j}] = \frac{1}{2}\{S_jS^{\dg}_{j},\rho\}-S_{j}^{\dg}\rho S_{j}$,
associated with the hole-spin relaxation strengths
of $\gamma_{j}$ and $\tilde{\gamma}_{j}$, respectively.
In the following numerical studies
we assume an identical strength of
$\gamma_{j}=\tilde{\gamma}_{j}\equiv \gamma_0 = 0.1\,\Omega$.
This choice assumes a similar timescale for the hole-spin
relaxation and the laser-induced Rabi oscillation period of the electron spin.
It thus allows us to demonstrate in a transparent
manner the effect of hole-spin
relaxation on optical manipulation.
Also, this choice is qualitatively consistent
with the experiment in Ref.\ \onlinecite{Gup01a}.

In the presence of hole-spin relaxation, the restricted subspace $\textbf{M}^{coh}$
constructed for the coherent evolution is no longer complete.
More specifically, the relaxation operators $S_j$ and $S^{\dg}_j$ ($j=1,\ldots 6$)
will cause a leakage of state from the subspace $\textbf{M}^{coh}$.
By applying the relaxation operators to the basis states of subspace
$\textbf{M}^{coh}$,
eight new states outside the subspace $\textbf{M}^{coh}$ appear.
We now denote the expanded Hilbert space by
$\textbf{M}^{inc}=\{|\psi_{i}\rangle, i=0,1,\tilde{1},2,3,4,5;
|\tilde{\psi}_{j}\rangle, j=1,2,...8\}$, which is complete for the evolution
of \Eq{Lindblad}.
Using the Lindblad equation (\ref{Lindblad}), in the following we study
the hole-spin relaxation effect for two possible situations.

{\it Instant manipulation}.---
In this case the spin manipulation is performed
right after the generation of the electron-hole pair.
Figure 3(a) shows the effect of hole-spin relaxation on the Rabi oscillation
of CB electron spin, where we have classified the spin-up and spin-down states
according to the CB electron spin,
i.e., summing the probabilities of states with the same CB spin state,
regardless of the hole spin states of the valence band.
The result in Fig.\ 3(a) shows that the hole-spin relaxation
will significantly influence the optical manipulation of the CB electron spin,
degrading it out from the desired coherent subspace
(and into the incoherent regime).
Also, differing from the conventional decoherence effect on {\it real}
two-state Rabi oscillation, besides the decay of the oscillation amplitudes
in Fig.\ 3(a), the sum of the probabilities of spin-up and
spin-down states eventually deviates from unity.
This deviation is caused by the relaxation-induced
real occupation of the intermediate states,
which are ruled out from the CB electron spin-up and spin-down states.
In the above calculation, we have adopted parameters such that
the Rabi oscillation of the CB spin has
a period similar to the hole-spin relaxation time.
By increasing the laser-system coupling strength,
more ``coherent'' oscillations can be observed.
Nevertheless, the present result implies that
 coherent optical manipulation in the {\it undoped} quantum well is
limited by the hole-spin relaxation, which is
typically on a timescale of picoseconds.


{\it Manipulation in the presence of hole-spin pre-relaxation}.---
In practice (e.g. in the experiment of Ref.\ \onlinecite{Gup01a}),
the manipulating laser pulse may be switched on with a certain delay time
after the resonant excitation pulse that prepared
the initially coherent electron-hole pair.
It is thus meaningful to show how the {\it pre-relaxation} of the
hole spin affects the subsequent manipulation of the electron spin.
Before the manipulating laser is applied,
the fast hole-spin relaxation may have led the valence band to a mixed
hole state,
while the CB electron spin that has not
yet coupled to the VB state
remains in the initially prepared pure state.
As the manipulating laser field is applied,
 Rabi coupling between the CB pure state and
the VB mixed state is established,
and relaxations among several involved
multi-particle states occur.
The optical manipulation in the presence of the above pre-relaxation
can be described by using the time-dependent Hamiltonian
$H=H_{0}+\Theta(t-\tau)H_{int}(t)$
for the same Lindblad master equation (\ref{Lindblad}).
Here $H_0$ is the field-free system Hamiltonian,
while $\Theta(t-\tau)H_{int}(t)$,
with $\Theta(t-\tau)$ being the step-function,
describes the system's interaction with
the time $\tau$-delayed manipulating laser field.

Shown in Fig.\ 3(b) and (c) are the
resulting Rabi oscillations of the CB electron spin
evaluated with two delay time values .
Compared to the instant manipulation counterpart in
Fig.\ 3(a), the results shown here indicate that
the pre-relaxation of the VB hole spin does not
significantly affect the manipulation
of the CB electron spin.
Intuitively, since the VB had been in a mixed state,
one would not expect any ``coherent'' signal
to appear in response to the optical manipulation,
but some coherent signal can be seen in the incoherent regime of Fig.\ 3(a).
\begin{figure}\label{Fig4}
\includegraphics*[scale=0.5,angle=0.]{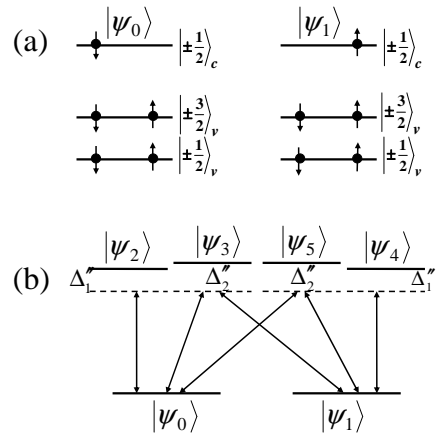}
\caption{ Diagram of band-edge states of a doped quantum well. (a)
Two degenerate state configurations with different electron spin in
the conduction band, between which quantum coherent rotation is to
be performed. (b) Intermediate states which mediate transition
between $|\psi_0\ra$ and $|\psi_1\ra$ during the optical
manipulation. Here the detunings $\Delta''_1$ and $\Delta''_2$
differ in a certain sense from their counterparts $\Delta'_1$ and
$\Delta'_2$ in the undoped quantum well, i.e.,
$\Delta''_1=\Delta_1+U_{eX}$, and $\Delta''_2=\Delta_2+U_{eX}$, with
$U_{eX}$ the Coulomb interaction energy of the doped electron and
the (virtually) generated exciton. }
\end{figure}
This counterintuitive result may be understood by noting that
in Fig.\ 3(a) the stationary mixed state is achieved
in the presence of the laser pulse.
In that completely mixed state, the CB electron spin jumps incoherently
between spin-up and spin-down states, and the hole state is not at all the
pre-relaxed valence band state achieved at the end of the first stage
in Fig.\ 3(b) and (c).
Thus, the laser pulse can still drive
coherent oscillation of the CB electron spin as in Fig.\ 3(a).
As a matter of fact, the analysis presented here
corresponds to the experiment in Ref.\ \onlinecite{Gup01a},
where the tipping laser pulse acted on
a pre-relaxed state.  The experimental result showed
that the electron spin can still be manipulated via
the intermediate mixed valence band
state by the laser pulse, which is consistent with our present understanding.

\section{ Doped Semiconductors}

So far our discussion has focused on the undoped
semiconductor quantum well,
where the excitation pump field
creates an electron in the CB while leaving a hole in the VB.
In this section we shall compare it with
the doped quantum well, in which
an excess electron is injected into the conduction band, and
the valence band remains fully occupied.
Shown in Fig.\ 4(a) is the level diagram
of a doped semiconductor quantum well.
Compared with the undoped system in Fig.\ 1(a), no hole exists
in the valence band of the doped system.
However, an electron in the VB can be
virtually excited to the CB during the manipulation.
This process is described by a number
of intermediate states shown in Fig.\ 4(b).
Using the same notation as for the undoped system, we denote the initial state
as $|\psi_{0}\rangle\equiv|1,0;1,1,1,1\rangle$,
the (conduction band) spin-flipped state as
$|\psi_{1}\rangle\equiv|0,1;1,1,1,1\rangle$,
and the relevant virtually excited intermediate states as
$|\psi_{2}\rangle\equiv|1,1;1,0,1,1\rangle$,
$|\psi_{3}\rangle\equiv|1,1;1,1,1,0\rangle$,
$|\psi_{4}\rangle\equiv|1,1;1,1,0,1\rangle$, and
$|\psi_{5}\rangle\equiv|1,1;0,1,1,1\rangle$.
\begin{figure}\label{Fig5}
\includegraphics*[scale=0.45,angle=0.]{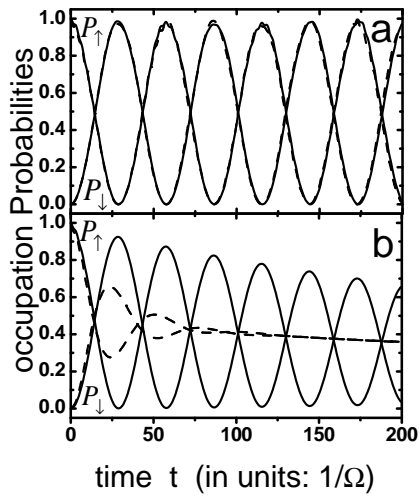}
\caption{Comparison of optical manipulation of the conduction-band
electron spin in doped and undoped quantum wells, in the (a) absence
and (b) presence of hole-spin relaxation. The solid and dashed
curves correspond to, respectively, results of the doped and undoped
quantum wells.}
\end{figure}
With the identification of these intermediate states,
the same analysis as for the undoped system can be carried out for both
the coherent manipulation and the hole-spin relaxation effect.
In the following calculation, we assume $U_{XX}\simeq U_{eX}$.
Any possible difference between them would be minor
in comparison with the large detuning energy,
resulting only in a negligibly small change of the
Rabi oscillation frequency.
As shown in Fig.\ 5(a), the doped and undoped systems have almost identical
responses to optical manipulation in the absence of hole-spin relaxation.
However, in the presence of hole-spin relaxation, the doped and undoped systems
behave very differently. As shown in Fig.\ 5(b),
while coherent manipulation in undoped semiconductor quantum wells
is sensitively limited by the hole-spin relaxation that virtually exists in the
intermediate states, optical manipulation of the doped system is much
more robust.
This result indicates that satisfactory coherent manipulation should be
achievable in doped systems.

\section{Concluding Remarks}

To summarize, we have presented a theoretical study of the optical manipulation
of electron spins in semiconductor quantum wells.
Our results provide not only a qualitative account for
the experimental findings in Ref.\ \onlinecite{Gup01a},
but also a considerably improved effective magnetic field
description based on the present many-particle-state STIRAP approach.
Moreover, we have also analyzed the effect of hole-spin relaxation,
and showed that while coherent optical control in undoped quantum wells
deteriorates significantly in the presence of
relatively fast hole-spin relaxation,
coherent control in doped systems can be quite robust against decoherence.

 We would also like to mention that
similar analysis based on the present formalism
can be straightforwardly applied to
the optical manipulation of spins in quantum dots,
a topic of great interest in recent years
\cite{Los00,Los03,Fol03,Pry02,Sha03,Los98,Ima99,Los99,Pri98,Los01,Zol01,Fen03,Sol03}.
In a quantum dot, the energy levels are discrete,
and thus long spin relaxation
time are anticipated.
%
In fact it has been shown experimentally that in quantum dots
both the electron and hole spins are almost frozen within the electron-hole
recombination timescale \cite{Pai01}, which is on the order of nanoseconds or longer.
Recent calculations predicted that for a typical quantum dot
the conduction electron spin relaxation
time is about $10^{-6}\sim 10^{-4}$ seconds \cite{Kha01},
and the hole-spin relaxation time is of order $10^{-8}$ seconds or longer \cite{woo04}.
These results suggest that the coherent optical manipulation
of electron spins discussed in this work can readily be carried out
in quantum dots.
Further, based on the result presented in Sec.\ IV,
almost identical control quality within the coherent time scale
is expected for either doped or undoped systems.
The latter is in general more experimentally accessible.
With optical means, the coherent conduction electron spin rotation
can be performed thousands of times
before the electron-hole recombination.

Finally, we remark that the present work has focused on quantum wells.
Therefore, the applicability of the discrete-level model
should be restricted to the optical manipulation of band-edge states.
Also, in our treatment the electron-hole Coulomb interaction
(exciton effect) is only accounted for in terms of some
phenomenological binding-energy parameters.
Although such treatment is typical for the
optical properties of quantum wells,
a more sophisticated treatment based on certain advanced techniques of
many-body theory might be desirable
in future work.
Also, in the study of the hole-spin relaxation effect,
the relaxation strengths have been put only by hand.
To make the study more realistic,
it would be desirable to present
calculations of the scattering rates between
various states based on the microscopic mechanisms.

\vspace{5ex}
{\it Acknowledgments.}
We sincerely thank the generous help from Professor YiJing Yan
in finalizing the manuscript.
Support from the Major State Basic Research Project No. G001CB3095 of China,
the Special Fund for ``100 Person Project" from Chinese Academy of Sciences,
and the National Natural Science Foundation of China
is gratefully acknowledged.


\begin{references}
\bibitem{note-zx}
Here the ``$z$"-representation means the combination of
$\sigma_z$-representation for electron spin, and $z$-axis as the
reference direction for the orbital angular momentum. Similarly, the
``$x$"-representation will be used later.
\bibitem{Not-1}
Two more considerations are involved in determination of the STIRAP
intermediate states: (i) The laser can only cause vertical
excitation of virtual electron-hole pair, because of the negligibly
small photon momentum. (ii) The intermediate state associated with
virtual excitation of electron-hole pair away from the band edge has
no effect on the STIRAP process considered here, since it does not
connect the initial state with the final state as shown in Fig.\
1(a).
\end{references}
\end{document}